\documentstyle[multicol,prl,aps,amssymb,graphicx]{revtex}
\setkeys{Gin}{width=0.9\linewidth}

\begin{document}
\draft

\title{Lower critical field \boldmath$H_{c1}$ and barriers for vortex 
       entry in Bi$_2$Sr$_2$CaCu$_2$O$_{8+\delta}$ crystals}

\author{M.~Nider\"ost$^1$, R.~Frassanito$^1$, M.~Saalfrank$^1$, A.C.~Mota$^1$,
        G.~Blatter$^2$, V.N.~Zavaritsky$^3$, T.W.~Li$^4$, and P.H.~Kes$^4$}

\address{$^1$Laboratorium f\"ur Festk\"orperphysik, ETH Z\"urich,
         8093 Z\"urich, Switzerland\\
         $^2$Theoretische Physik, ETH Z\"urich, 8093 Z\"urich, 
         Switzerland\\ 
         $^3$Kapitza Institute for Physical Problems, 117334 Moscow,
         Russia\\
         $^4$Kamerlingh Onnes Laboratorium, Rijksuniversiteit Leiden, PO 
         Box 9506, 2300 RA Leiden, The Netherlands}

\date{\today}
\maketitle

\begin{abstract} 

The penetration field~$H_{p}$ of $\mathrm{Bi_2Sr_2CaCu_2O_{8+\delta}}$ 
crystals is determined from magnetization curves for different field 
sweep rates~$dH/dt$ and temperatures.  The obtained results are 
consistent with theoretical reports in the literature about vortex 
creep over surface and geometrical barriers.  The frequently observed 
low-temperature upturn of~$H_{p}$ is shown to be related to metastable 
configurations due to barriers for vortex entry.  Data of the true 
lower critical field~$H_{c1}$ are presented.  The low-temperature 
dependence of $H_{c1}$ is consistent with a superconducting state with 
nodes in the gap function.  [PACS numbers: 74.25.Bt, 74.60.Ec, 
74.60.Ge, 74.72.Hs]
\end{abstract}


\begin{multicols}{2}
\narrowtext
High-temperature superconductors (HTSC's) are strong\-ly type~II and as 
such their $H-T$ phase diagram involves complete flux expulsion 
(Meissner state) at low fields and magnetic field penetration in the 
form of quantized flux-lines or vortices (mixed state) at higher 
fields.  The onset of the mixed state, where vortex penetration 
becomes energetically favorable, is defined as the lower critical 
field $H_{c1} \propto 1/\lambda^2$ (here, $\lambda$ is the London 
penetration depth).  From the temperature dependence of~$H_{c1}$, 
important information can be gained, particularly, regarding the 
symmetry of the superconducting state, since the appearance of gap 
nodes strongly modifies the $T$-dependence of the superfluid density 
and thereby the penetration depth~$\lambda(T)$.

The experimental determination of~$H_{c1}$ has been a challenging and 
controversial problem since the beginning of HTSC research.  Not only 
the values reported for the first flux penetration field~$H_{p}$ are 
scattered over a wide field range, but in strongly layered 
superconductors with~$H$ perpendicular to the planes, striking 
features have been observed, such as a marked upturn of~$H_{p}$ for 
temperatures $T \lesssim T_{c}/2$ 
\cite{kopylov90,zava91,mota91,chiku1,metlushko93,zeldov95}.  The 
origin of the positive curvature of~$H_{p}$ at low temperatures has 
been the matter of various speculations.  Different mechanisms have 
been proposed: Bean-Livingston surface 
barriers~\cite{kopylov90,mota91,chiku1,zeldov95}, bulk 
pinning~\cite{zava91,zeldov95}, a modification of the character of the 
field penetration in layered structures~\cite{metlushko93}, a 
low-temperature enhancement of the superconducting order parameter in 
the normal layers~\cite{koyama90}, etc.  

For type~II superconductors, there are at least two kinds of barriers 
which hinder the system from reaching a thermodynamic equilibrium 
state: (i)~surface and geometrical barriers~\cite{bl,indemb,zeldov94} 
governing vortex penetration into the superconductor, (ii)~bulk 
pinning barriers governing vortex motion in the superconductor.  For 
the determination of $H_{c1}$, the relevant barriers are those which 
govern vortex entry into the sample.  Of particular interest in this 
context is the phenomenon of vortex creep over surface and geometrical 
barriers.  From the theoretical point of view this subject has been 
discussed in Refs.\ \onlinecite{zeldov94} and \onlinecite{burl}, 
however, to our knowledge no systematic experimental study has been 
carried out so far.

In this letter we investigate the dependence of~$H_{p}$ on the 
magnetic field sweep rate~$dH/dt$ of isothermal magnetization curves 
for the strongly layered $\mathrm{Bi_2Sr_2CaCu_2O_{8+\delta}}$ 
(Bi2212) material.  This has been done for crystals with different 
cross-sections along the $c$-axis.  For specimens with ellipsoidal 
cross-sections geometrical barriers can be 
neglected~\cite{zeldov94,majer95}, so that the relevant barriers for 
vortex entry are expected to be surface barriers.  Indeed, for our 
specimen with an ellipsoidal-like cross-section, the~$H_{p}$ 
vs.~$dH/dt$ dependence obtained at high sweep rates is well described 
by vortex creep over surface barriers~\cite{burl}.  However, for 
decreasing $dH/dt$~rates, these barriers are observed to collapse.  
The subsequent saturation of~$H_{p}$ at lower rates indicates that the 
system is in equilibrium with respect to surface barriers.  This is 
interpreted as strong evidence that we have reached the lower critical 
field~$H_{c1}$ in our measurements.  The obtained low-temperature 
dependence of~$H_{c1}$ is in good agreement with recent microwave 
absorption measurements of the penetration 
depth~$\lambda$~\cite{jacobs,Lee96}.  On the other hand, for specimens 
with rectangular cross-sections, no significant vortex creep over the 
relevant barriers for vortex entry could be observed, in consistency 
with theoretical results for geometrical barriers~\cite{zeldov94}.  
The low-temperature upturns of $H_{p}$ observed for specimens of 
either cross-section are then explained in terms of measurements where 
the system is out of equilibrium with respect to barriers for vortex 
entry.

\begin{figure}
   \includegraphics{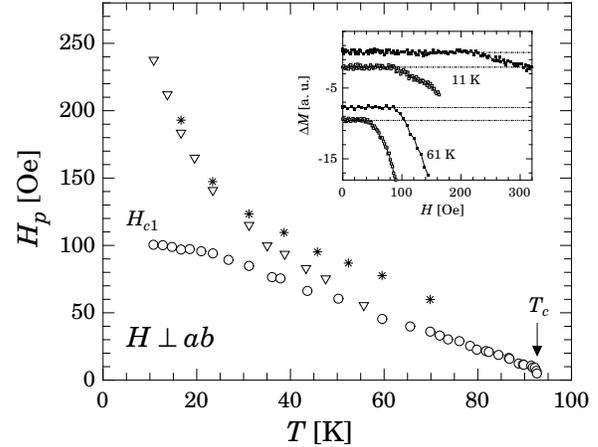}
  \caption{Temperature dependence of the first flux penetration field~$H_{p}$ 
           of the specimen with an ellipsoidal-like cross-section 
           for different field sweep rates: (\/{\large 
		   $\circ$})~$dH/dt \leq 1 \cdot 10^{-4}\, \mathrm{Oe/s}$, (\/{\large 
		   $\ast$})~$dH/dt \geq 5 \cdot 10^{-2}\, \mathrm{Oe/s}$, and 
		   (\/$\triangledown$)~$dH/dt = 8 \cdot 10^{-3}\, \mathrm{Oe/s}$. 
		   The inset shows the deviation from the Meissner slope for $T = 11 \, 
		   {\mathrm{K}}$ and $61 \, {\mathrm{K}}$.  The upper curves 
		   (\/{\scriptsize $\blacksquare$}) are measured at a
		   rate $dH/dt \geq 5 \cdot 10^{-2}\, \mathrm{Oe/s}$ and the 
		   lower ones (\/{\scriptsize $\square$}) at 
		   $dH/dt \leq 1 \cdot 10^{-4}\, \mathrm{Oe/s}$.  The data are 
		   scaled with the demagnetization factor $N = 0.96 (\pm15\%)$.}
  \label{Hp(T)}
\end{figure}

The crystals under investigation have been grown with different 
techniques and have different shapes.  The specimen with an 
ellipsoidal-like cross-section along the $c$-axis has been grown in a 
$\mathrm{ZrO_2}$ crucible~\cite{zavaritzky} and is approximately 
$1\times1.3 \times0.05~{\rm mm}^3$ in size.  The specimens with 
rectangular cross-sections along the $c$-axis have been grown with the 
traveling solvent floating zone method~\cite{Li} and are slightly 
larger in size.  The experiment is performed in a non-commercial SQUID 
magnetometer where the sample is stationary in the pick-up coil.  The 
field~$H$ is supplied by a superconducting coil working in a 
non-persistent mode.  For the magnetization curve measurements, the 
sample is zero field cooled from above~$T_{c}$ and stabilized at a 
fixed temperature (the residual field of the cryostat is $<\,10 \, 
{\rm mOe}$).  Further, the field~$H$ is applied at a fixed rate 
$dH/dt$.  For~$H_{p}$, we select the field at which a deviation from 
Meissner shielding occurs (see inset of Fig.\,\ref{Hp(T)}).
 
\begin{figure}
   \includegraphics{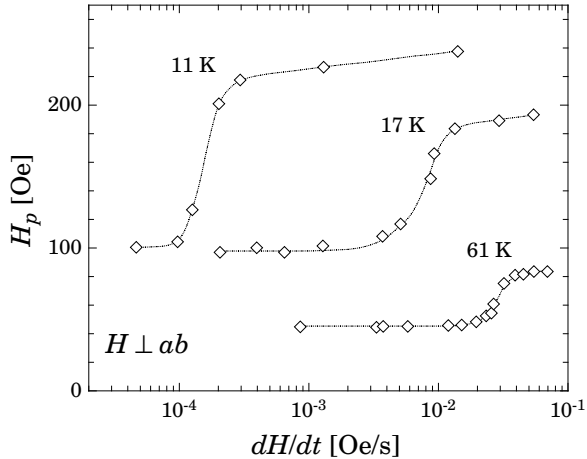}
   \caption{For the specimen with an ellipsoidal-like cross-section, the  
            first flux penetration field~$H_{p}$ vs.~the applied magnetic 
            field sweep rate $dH/dt$ is displayed for different temperatures.
            The dotted lines are guides to the eyes.  The data are scaled 
            with the demagnetization factor $N = 0.96 (\pm15\%)$.}
   \protect\label{Hp(dH/dt)}
\end{figure}

Fig.\,\ref{Hp(dH/dt)} shows the dependence of~$H_{p}$ on the field 
sweep rate~$dH/dt$ at three characteristic temperatures for the 
specimen with an ellipsoidal-like cross-section.  A~sharp step in 
the~$H_{p}$ vs.~$dH/dt$ curves is observed at all three temperatures.  
At high sweep rates, the curves display a finite slope which is most 
pronounced at $T = 11\,{\mathrm{K}}$ and $17\,{\mathrm{K}}$ and to a 
lesser degree at $T = 61\,{\mathrm{K}}$.  These slopes are in good 
agreement with the theoretical results of Burlachkov {\it 
et~al.}~\cite{burl} for vortex creep over surface barriers (see the 
analysis below).  Proceeding from the creep regime towards lower 
rates, the sharp drop signals a collapse of the surface barriers (or 
equivalently a heat pulse) which leads to a rapid flux entry, possibly 
in terms of flux jumps.  Indeed, an avalanche type of flux penetration 
into a type~II superconductor has been reported by Dur\'an {\it et 
al.}~\cite{duran}.  On decreasing the sweep rates even further, the 
flux penetrates into the sample through the occurrence of rare events 
when the applied magnetic field reaches~$H_{c1}$.  We thus identify 
the saturated value of~$H_{p}$ with the lower critical field~$H_{c1}$ 
and use it to obtain information about the nature of the 
superconducting state.

Analogous measurements have been done on thin specimens with 
rectangular cross-sections.  According to Zeldov~\textit{et 
al.}~\cite{zeldov94}, for such specimens, geometrical barriers build 
up over a~distance $s$ from the sample edges, where~$s$ is the sample 
thickness.  The energy required to overcome such~a barrier is 
macroscopic~$\sim\varepsilon_{0}s$ so that vortex creep over 
geometrical barriers is expected to be very weak (here $\varepsilon_{0} 
= (\Phi_{0} / 4 \pi \lambda)^2$ and $\Phi_{0}$ is the unit flux).  As a 
matter of fact, we were not able to detect significant vortex creep 
over geometrical barriers within the experimental time scale $10^{-5} 
\lesssim dH/dt \lesssim 10^{-1} \, {\mathrm{Oe/s}}$.

We proceed with a brief discussion of vortex creep over surface 
barriers.  As shown by Burlachkov {\it et al.}~\cite{burl}, the 
surface barrier for pancake-vortices is given by $U \simeq 
\varepsilon_{0}d \ln{(0.76 H_{c}/H)}$, where~$H_{c}$ is the 
thermodynamic critical field and $d$ the interlayer distance.  During 
the time~$t$, thermal creep allows to overcome barriers of size $U(t) 
\simeq T \ln(t/t_{0})$, where~$t_{0}$ is a ``microscopic'' time 
scale~\cite{blatter}.  Equating the two expressions one obtains the 
time dependence of the penetration field for the pancake-vortex 
regime,
\begin{equation} \label{Hp1}
  H_{p}(t) \simeq H_c \,(t/t_{0})^{-T/\varepsilon_{0} d} \,.
\end{equation}
With the definition $T_{0}=\varepsilon_{0}d /\ln{(t/t_{0})}$ 
Eq.~(\ref{Hp1}) becomes $H_{p} \simeq H_{c} \exp{(-T/T_{0})}$.  At 
higher temperatures creep proceeds via half loop excitations of 
vortex-lines~\cite{burl}.  The creation of a half loop saddle 
configuration involves an energy $U \simeq 
\pi\varepsilon\varepsilon_{0}^2 c \,\ln^2{(j_{0}/j)}/2 \Phi_{0} j$, 
where $j_{0}$ is the depairing current density and~$\varepsilon 
=(m/M)^{1/2}<1$ the anisotropy parameter.  With a similar analysis as 
above, the time dependence of $H_{p}$ for the vortex-line regime takes 
the~form
\begin{equation} \label{Hp2}
  H_{p}(t) \approx H_c \,\frac{\pi}{2\sqrt{2}}\, 
  \frac{\varepsilon\varepsilon_{0} \xi}{T\ln{t/t_{0}}} \, 
  \ln^2{\left(\frac {H_{c}} {H_{p}}\right)}\,,
\end{equation}
where~$\xi$ is the coherence length.  According to Ref.\,\cite{burl}, 
this expression gives a temperature dependence $H_{p} \propto (T_{c} 
-T)^{3/2}/T$.  The crossover between the two regimes occurs when the 
size of the half loop excitation is of the order of the interlayer 
distance~$d$.  This is expected to occur at a temperature $T^{*} 
\approx T_{0}(T^{*}) \ln{(d/\varepsilon\xi)}$.

\begin{figure}
  \includegraphics{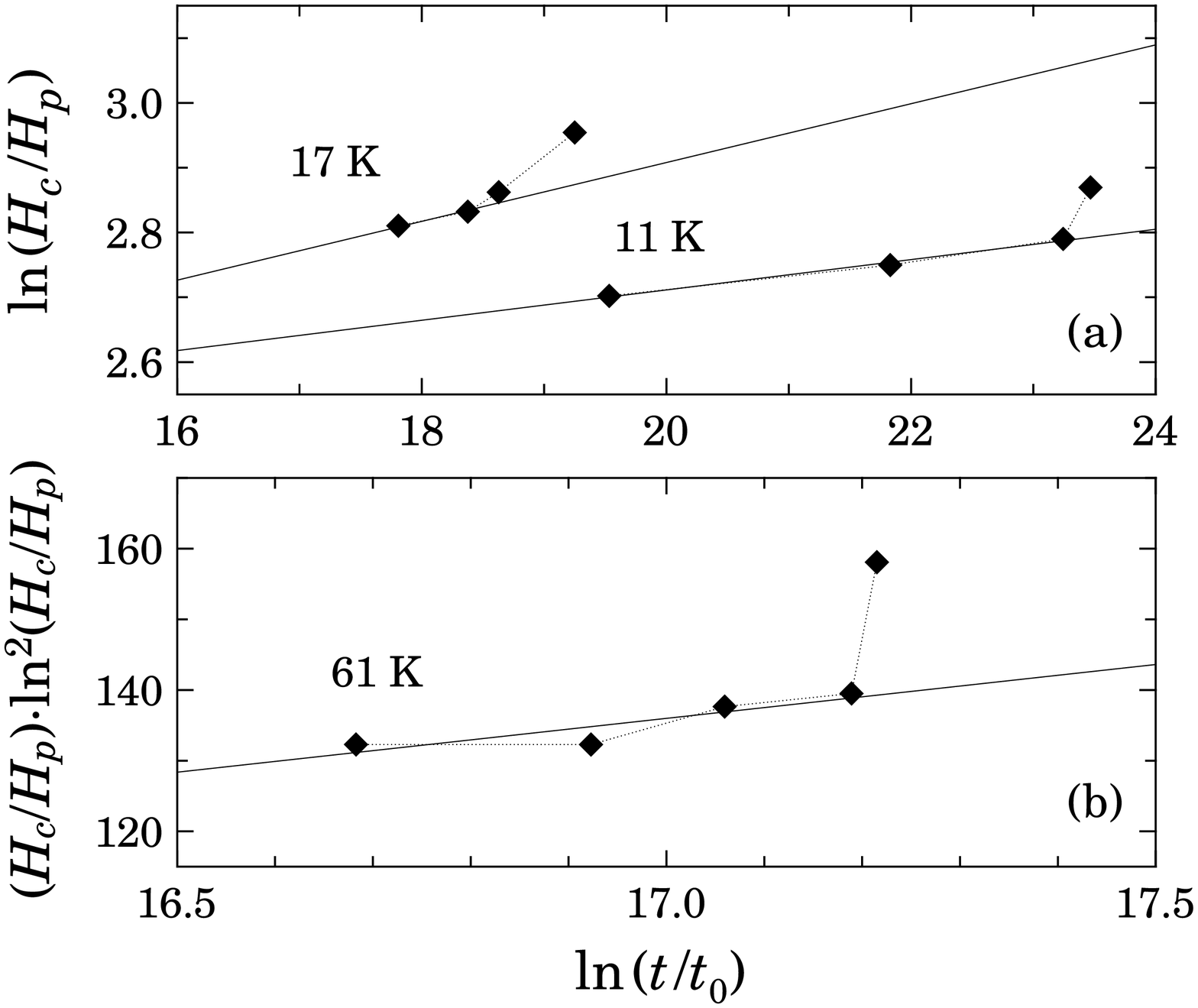}
  \caption{(a) $T = 11\,{\mathrm{K}}$ and $17\, {\mathrm{K}}$ data of 
           the upper plateau in Fig.\,\ref{Hp(dH/dt)} in a~different
           representation.  Here we choose the time origin $t = 0$ at 
           $H = H_{c1}$. The critical field $H_{c}(T) \simeq H_{c}(0)\,
           (1-T/T_{c})$ is calculated with $\lambda_{ab}(0)$ as obtained 
           from the~$H_{c1}$ data in Fig.\,\ref{Hp(T)} and the coherence
           length $\xi(0) \simeq 25\,{\mathrm{\AA}}$.
		   We assumed $t_{0} \simeq 10^{-6}\, {\mathrm{s}}$.
           (b) $T = 61\, {\mathrm{K}}$ data at high sweep rates in
           Fig.\,\ref{Hp(dH/dt)} in a convenient representation. The lines 
           in Figs.\,\ref{ln(t/to)}(a) and (b) are fits according to
           Eqs.\,(\ref{Hp1}) and (\ref{Hp2}), respectively. }
  \label{ln(t/to)}
\end{figure}

According to (\ref{Hp1}), creep of pancake-vortices over surface 
barriers results in a linear dependence of $\ln(H_{c}/H_{p})$ on 
$\ln(t/t_{0})$ with a slope $T/\varepsilon_{0}d$.  This dependence is 
given in Fig.\,\ref{ln(t/to)}(a) using the data at high cycling rates 
in Fig.\,\ref{Hp(dH/dt)}.  For $T\,=\,11\,{\mathrm{K}}$ and 
$17\,{\mathrm{K}}$, the values of $T/\varepsilon_{0}d$ obtained from 
the fits are in good agreement with the values estimated from the 
penetration depth~$\lambda_{ab}(T)$ (see Table \,\ref{FitPara1}). \\

\begin{table}
\caption{$T/\varepsilon_{0}d$ values calculated with the penetration depth 
         $\lambda_{ab}(T)$ as obtained with formula $H_{c1} = (\Phi_{0}/4 \pi  
         \lambda_{ab}^{2}) \cdot \ln \kappa$ from the $H_{c1}$ data 
         in Fig.\,\ref{Hp(T)} and values of $T/\varepsilon_{0}d$ obtained 
         from the fits in Fig.\,\ref{ln(t/to)}(a).}
		 \protect\label{FitPara1} 
  \begin{center}
	  \begin{tabular}{l|cc}
		 & $T = 11\,{\mathrm{K}}$ & $T = 17\,{\mathrm{K}}$ \\
		 \hline
		 $T/\varepsilon_{0}d$ (calculated) & 1/34 & 1/21 \\
		 $T/\varepsilon_{0}d$ (from fit) & 1/43 $\pm 30\%$ & 1/22 $\pm 30\%$\\
	  \end{tabular}
  \end{center}
\end{table}

\noindent On the other hand, for $T\,=\,61\,{\mathrm{K}}$ the above 
analysis for pancake-vortices is not satisfactory since the 
$T/\varepsilon_{0}d$ value obtained from the fit is one order of 
magnitude smaller than the calculated value.  According to 
(\ref{Hp2}), for creep of vortex-lines over surface barriers 
$(H_{c}/H_{p})\cdot\ln^{2}(H_{c}/H_{p})$ vs.  $\ln(t/t_{0})$ is linear 
with a slope $2 \sqrt{2} T/\pi \varepsilon \varepsilon_{0} \xi$.  For 
the $T\,=\,61\,{\mathrm{K}}$ data at high cycling rates in 
Fig.\,\ref{Hp(dH/dt)}, such a representation is given in 
Fig.\,\ref{ln(t/to)}(b).  The slope $2 \sqrt{2} T/\pi \varepsilon 
\varepsilon_{0} \xi$ obtained from the fit is $15\, \pm\,3$, in good 
agreement with the value~$16$, estimated with the help of the GL 
formula for the coherence length $\xi(T)$, with 
$\xi(0)\,\simeq\,25\,{\mathrm{\AA}}$ and an anisotropy parameter 
$\varepsilon\,\simeq\,1/100$.  From these considerations we conclude 
that the behavior of~$H_{p}$ at high sweep rates is determined by 
creep of pancake-vortices (at low temperatures) and vortex-halfloops 
(at higher temperatures) over surface barriers.  The estimated value 
of the activation barrier for vortex entry is $U\,\simeq\,300 \, 
{\mathrm{K}}$ for temperatures~$T$ between 10 and $20 \, {\mathrm{K}}$ 
and $U\,\simeq\,1200 \, {\mathrm{K}}$ for~$T\,\simeq\,60 \, 
{\mathrm{K}}$ (here we set the time origin $t\,=\,0$ when the applied 
field $H$ reaches~$H_{c1}$ and assume $t_{0}\,\simeq\,10^{-6}\, 
\mathrm{s}$).

In Fig.\,\ref{Hp(T)} we make use of the curves in 
Fig.\,\ref{Hp(dH/dt)} to determine $H_{p}(T)$ with three different 
criteria: (i)~$H_{p}$ data ({\large $\circ$}) are taken at low sweep 
rates so as to guarantee that they lie in the regime where the system 
is in equilibrium with respect to surface barriers (these data 
represent~$H_{c1}(T)$), (ii)~$H_{p}$ data ({\large $\ast$}) are taken 
at high rates so that they always lie in the creep regime, and 
(iii)~the field~$H$ is swept at the constant rate $dH/dt = 
8\,\cdot\,10^{-3}\, \mathrm{Oe/s}$ for all temperatures.  In this 
case, the values of~$H_{p}$ ($\triangledown$) lie on the 
$H_{c1}$-curve for $T \gtrsim 60 \, {\mathrm{K}}$, they go through the 
step-like crossover for temperatures $30 \, {\mathrm{K}} \lesssim T 
\lesssim 60 \, {\mathrm{K}}$, and finally, for $T \lesssim 30 \, 
{\mathrm{K}}$ they lie in the creep regime.  From a comparison of the 
curves in Fig.\,\ref{Hp(T)}, it follows that the low-temperature 
upturn of~$H_{p}$ results from measurements where the system is out of 
equilibrium with respect to surface barriers.

As shown in Fig.\,\ref{Hp(T)}, for increasing temperatures the 
$T$-dependence of~$H_{p}$ in the creep regime, ({\large $\ast$}) and 
($\triangledown$) for $T \lesssim 30 \, {\mathrm{K}}$ and ({\large 
$\ast$}) for $T \gtrsim 30 \, {\mathrm{K}}$, undergoes a crossover 
from an exponential behavior to a weak power-law.  This is in 
agreement with the results presented above and indicates that the 
crossover from creep of pancake-vortices to creep of vortex-halfloops 
over surface barriers takes place at $T^{*} \approx 30-40\,{\mathrm{K}}$ 
(in consistency with the estimate $T^{*} \approx T_{0}(T^{*}) 
\ln{(d/\varepsilon\xi)} \approx 40 \, {\mathrm{K}}$ obtained using the 
parameters for Bi2212 and $\ln{t/t_{0}} \approx 20$).

Making use of the correct~$H_{c1}$ data for~$H \perp ab$, ({\large 
$\circ$}) in Fig.\,\ref{Hp(T)}, we have investigated the temperature 
dependence of the penetration depth~$\lambda_{ab}$ using the 
expression $H_{c1} = (\Phi_{0}/4 \pi \lambda_{ab}^{2}) \cdot \ln 
\kappa$\/, (here we assume that $\ln \kappa$ is $T$-independent).  For 
$11 \, {\mathrm{K}} <T \lesssim 30 \, {\mathrm{K}}$, we find a linear 
behavior $\lambda_{ab}(T) -\lambda_{ab}(0) \propto T$ with slope 
$d\,\lambda/dT \simeq 10 \,{\mathrm{\AA/K}}$, where a linear 
extrapolation to the data gives a $T = 0$ penetration depth 
$\lambda_{ab}(0) \simeq 2700 \,{\mathrm{\AA}}$.  Our measurements thus 
confirm the linear low-temperature dependence of~$\lambda_{ab}$ as 
well as the slope $d\,\lambda/dT$, which have been recently determined 
with microwave absorption techniques on clean Bi2212 
crystals~\cite{jacobs,Lee96}.  A linear low-temperature dependence 
of~$\lambda_{ab}$, as first observed by Hardy {\it et 
al.}~\cite{hardy93} in YBCO crystals, is consistent with a 

\begin{figure}
  	\includegraphics{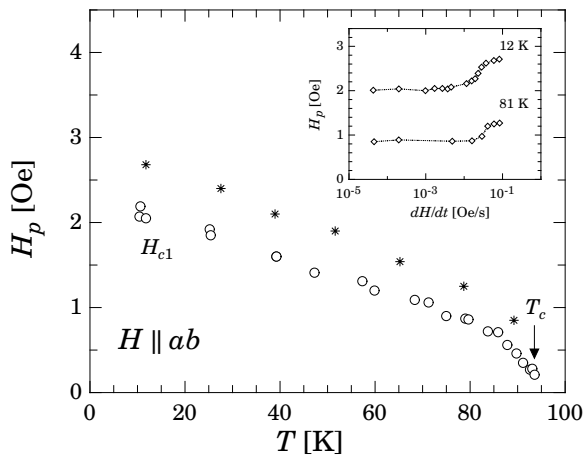}
  	\caption{Temperature dependence of~$H_{p}$ ($H \parallel ab$) 
  	         for a specimen with rectangular cross-section and size
  	         $2\times3.9\times0.05~{\rm mm}^3$ for different sweep rates: 
  	 	     ({\large $\ast$}) $dH/dt \geq 6 \cdot 10^{-2}\, 
  	 	     \mathrm{Oe/s}$ and ({\large $\circ$}) $dH/dt \leq 6 \cdot 
  	 	     10^{-3}\, \mathrm{Oe/s}$.  The inset shows the $H_{p}$~vs. $dH/dt$ 
  	 	     dependence at $T = 12 \, {\mathrm{K}}$ and $81 \,{\mathrm{K}}$.}
    \label{Hp||(T)}
\end{figure}
\noindent 

\noindent superconducting state with line nodes on the Fermi 
surface~\cite{annett}, e.g., a $d$-wave symmetry of the order 
parameter.

Similar investigations of~$H_{p}$ have been done for $H \parallel ab$.  
The misalignment angle between $H$ and the $ab$-planes has been 
estimated to be smaller than $2^{\small{\circ}}$ from measurements at 
magnetic fields in the range $H_{p\parallel} \ll H < H_{p\perp}$, (see 
Ref.~\cite{naka93}; here $H_{p\parallel}$ and $H_{p\perp}$ are the 
parallel and perpendicular penetration fields).  For this 
configuration geometrical barriers are irrelevant.  As shown in 
Fig.\,\ref{Hp||(T)}, for temperatures~$T$ between 10 and $70 \, 
{\mathrm{K}}$, $H_{c1}(T)$ has an~approximately linear behavior.  
According to Ref.~\cite{clem91b}, for temperatures not so close 
to~$T_{c}$, the correct description of~$H_{c1}$ in strongly layered 
materials is $H_{c1} = \Phi_{0}/(4 \pi \lambda_{ab} \lambda_{c}) \cdot 
[\ln(\lambda_{ab}/d) + 1.12]$.  With this formula we calculated the 
penetration depth in $c$-direction~$\lambda_{c}$ using the previously 
determined $\lambda_{ab}$ data.  For $T \lesssim 40 \, {\mathrm{K}}$, 
$\lambda_{c}$ is approximately linear in~$T$ with a slope of $0.1 
\,{\mathrm{\mu m/K}}$.  The extrapolated~$T = 0$ value is 
$\lambda_{c}(0) \simeq 15 \,{\mathrm{\mu m}}$.  Fig.\,\ref{Hp||(T)} 
further shows~$H_{p}$ data at high sweep rates ({\large $\ast$}).  
Contrary to the case for $H \perp ab$, no upturn of~$H_{p}$ is 
observed here for temperatures $T \lesssim T_{c}/2$.  We attribute 
this difference to the absence of the strong pancake vortex creep 
regime at low temperatures for $H \parallel ab$.  From the $H_{p}$~vs.  
$dH/dt$ dependence in the inset of Fig.\,\ref{Hp||(T)}, we obtain an 
activation barrier for vortex entry $U \simeq 200\, {\mathrm{K}}$ at 
$T = 12 \, {\mathrm{K}}$ and $U \simeq 1300\, {\mathrm{K}}$ at $T = 81 
\, {\mathrm{K}}$.

Close to~$T_{c}$ a~downward bending of~$H_{c1}$ is observed for $H 
\perp ab$ as well as $H \parallel ab$ (see Figs.\,\ref{Hp(T)} and 
\,\ref{Hp||(T)}).  This can be understood on the basis of an entropic 
downward renormalization of the vortex line free energy due to 
fluctuations of the order parameter around its mean-field 
form~\cite{blatter93b}.  The decrease in the free energy $f_{l} = 
\varepsilon_{l} - Ts_{l}$ then leads to a drop in~$H_{c1}$ as~$T$ 
approaches~$T_{c}$ (here, $\varepsilon_{l}$ is the line energy of the 
vortex excitation and~$s_{l}$ is the line entropy).

To conclude, isothermal magnetization curves were measured on Bi2212 
crystals for configurations of negligible geometrical barriers.  At 
very low sweep rates we found saturated values of the penetration 
field~$H_{p}$ which we interpret as the true lower critical 
field~$H_{c1}$.  The low-temperature dependence of~$H_{c1}$ is 
consistent with recent magnetic penetration depth 
measurements~\cite{jacobs,Lee96} suggesting a superconducting state 
with nodes in the gap function.  Based on the results obtained for 
Bi2212, we argue that the frequently reported low-temperature upturns 
of~$H_{p}$ in strongly layered superconductors with~$H$ perpendicular 
to the planes 
\cite{kopylov90,zava91,mota91,chiku1,metlushko93,zeldov95} can be 
explained in terms of measurements where the system is out of 
equilibrium with respect to barriers for vortex penetration.

It is a great pleasure to acknowledge K.~Aupke, C.~de~Morais-Smith, 
V.B.~Geshkenbein, A.~Suter, T.~Teruzzi, and P.~Visani for fruitful 
discussions and helpful contributions.  This work has been supported 
by the Swiss National Science Foundation.

\end{multicols}

\end{document}